# Assessing provincial carbon budgets for residential buildings to advance net-zero ambitions [*]

Hong Yuan [1], Minda Ma [2 *, 3], Nan Zhou [4], Zhili Ma [1], Chunbo Zhang [5]

1. School of Management Science and Real Estate, Chongqing University, Chongqing, 400045, PR China
2. School of Architecture and Urban Planning, Chongqing University, Chongqing, 400045, PR China
3. Building Technology and Urban Systems Division, Energy Technologies Area, Lawrence Berkeley National Laboratory, Berkeley, CA 94720, United States
4. Energy and Resources Group, University of California, Berkeley, CA 94720, United States
5. Department of Civil, Environmental and Geomatic Engineering, University College London, London, WC1E 6BT, UK

- Corresponding author: Prof. Dr. Minda Ma, Email: maminda@lbl.gov

  Homepage: https://globe2060.org/MindaMa/

[*] The co-authors declare that this manuscript was authored by an author at Lawrence Berkeley National Laboratory under Contract No. DE-AC02-05CH11231 with the U.S. Department of Energy.

## Graphical abstract

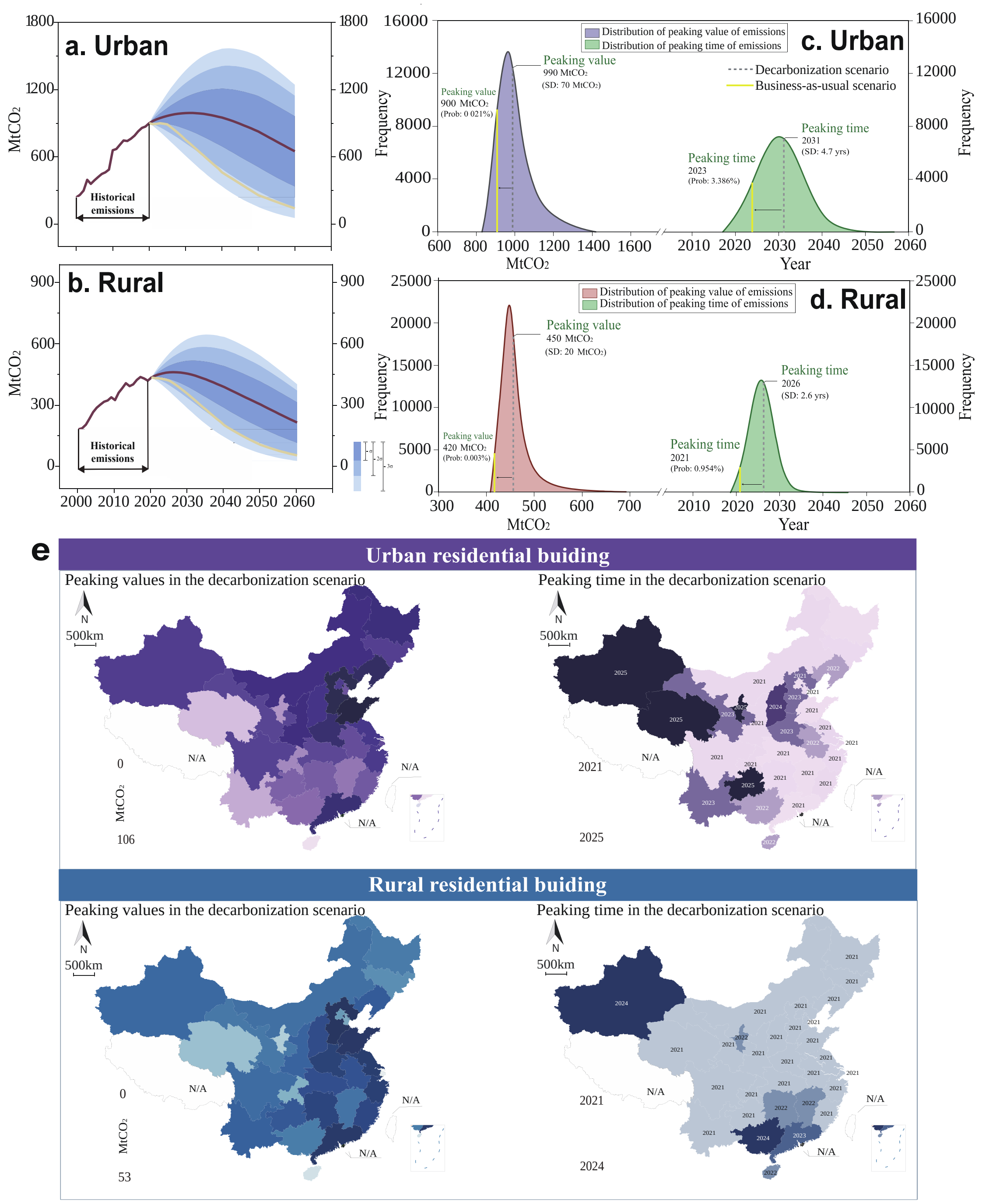

**Graphical abstract.** Static scenarios of carbon emissions for residential buildings and prospective ranges of dynamic simulations with different probabilities in (a) urban and (b) rural China, 2001-2060; distribution of peaking value and peaking time of (c) urban and (d) rural residential buildings; (e) peaking values and time in urban and rural China under the decarbonization scenario.

## Highlights

- Developed an assessment framework integrating static and dynamic modeling for carbon budgets.
- Provided a practical tool to generate robust provincial carbon budget estimates with limited data.
- Under the business-as-usual scenario, urban residential buildings will peak at 990 $MtCO_2$ in 2031.
- The urban residential peak decreases from 990 to 900 $MtCO_2$ under the decarbonization scenario.
- The Northwest holds the greatest reduction potential (38.14 $MtCO_2$) for urban and rural buildings.

## Abstract

Assessing provincial carbon budgets for residential building operations is a crucial strategy for advancing China's net-zero ambitions. This study develops an advanced assessment framework that integrates static and dynamic modeling to evaluate emission pathways and carbon budgets for residential buildings at the provincial level. The framework provides a practical tool for generating robust carbon budget estimates with limited data, thereby enhancing its applicability for policy-oriented planning. Findings show that (1) in the business-as-usual scenario, the emissions for urban and rural residential buildings are projected to peak at 990 (±0.7) and 450 (±0.2) mega-tons of carbon dioxide ($MtCO_2$), respectively, with peak years occurring in 2031 (±4.7) and 2026 (±2.6). (2) In the decarbonization scenario, peak emissions decrease to 900 $MtCO_2$ and 430 $MtCO_2$ for urban and rural buildings, respectively. (3) The provinces with the highest emission reduction requirements are Henan (16.74 $MtCO_2$), Xinjiang (12.59 $MtCO_2$), Gansu (9.87 $MtCO_2$), Hebei (8.46 $MtCO_2$), and Guangdong (3.37 $MtCO_2$), with Northwest China shouldering the greatest reduction responsibility, totaling 38.14 $MtCO_2$. In conclusion, this study provides a dynamically optimized carbon budget assessment tool for residential buildings, offering valuable insights for government policy-making and playing a key role in facilitating the low-carbon transition of China's building sector during the pre-2030 planning period, ultimately contributing to the goal of achieving net-zero emissions in the building sector by mid-century.

## Abbreviations and symbols

Please find this part in the Appendix B.

## 1. Introduction

The global building sector accounts for over one-third of both carbon dioxide ($CO_2$) emissions and energy consumption [1]. $CO_2$ emissions from building operations have increased by about 5% [approximately 10 giga-tons (Gt) of $CO_2$] since 2020, reaching historic highs [2]. Despite concerning levels of emissions in the building sector, the substantial potential for energy savings and emissions reduction still exists [3]. Specifically, the energy savings and emission-reduction potential of the building sector can reach up to nearly 74%, which is greater than that of the industrial and transportation sectors [4]. In 2020, China's building operations generated a total of 2.16 $GtCO_2$ in carbon emissions, accounting for 21% of global building carbon emissions (ranking first globally) [5]. Given the current trends, energy use and carbon emissions for the building sector in China are projected to continue rising [6]. Notably, residential buildings account for over 60% of energy use and nearly half of the emissions in the building sector [7]. Furthermore, recent evidence indicates that residential buildings have greater potential and cost-effectiveness for enhancing energy efficiency and reducing emissions than nonresidential buildings [8]. Hence, to achieve the net zero goal early, China should focus on decarbonizing residential buildings.

While existing studies have developed a framework for assessing energy use and emissions for buildings, aiming at the macro-national scale [9, 10]. It is worth noting that significant regional disparities in energy use and emissions exist across provinces in China's building sector, driven by variations in economic development [11], climatic conditions [12], energy structures [13], and policy implementation [14]. Therefore, developing a carbon budget assessment scheme across provinces is crucial. In particular, there is a relative paucity of studies on optimizing carbon peaks and formulating allocation schemes for residential buildings at the provincial level. To bridge these gaps, this study explores the following questions:

- What are the national trends in operational carbon emissions from residential buildings?
- How do the peaks of carbon emissions from building operations differ across provinces?
- How can the carbon budget be allocated to each province to maximize decarbonization potential?

To address these challenges, this study develops a carbon emission model for residential buildings in urban and rural China, grounded in the Kaya identity. Two static scenarios—business-as-usual and decarbonization—are used to analyze the emission peaks of residential buildings in China and each province through 2060. Monte Carlo simulation is then employed for a dynamic analysis of these two scenarios, predicting the emission-peaking probabilities for urban and rural residential buildings. Furthermore, an optimized assessment approach for provincial carbon budgets is introduced, targeting maximum emission reductions in support of the net-zero goal.

The key contribution of this study is the development of a practical tool for generating robust provincial carbon budget estimates for residential buildings under conditions of limited data. Methodologically, the study advances the field by integrating a static-dynamic modeling framework with optimization under net-zero targets, while explicitly capturing the distinct emission characteristics of urban and rural buildings and the heterogeneity across provinces. Specifically, the framework dynamically optimizes peaking values and peaking years for each province, ultimately producing a carbon budget allocation scheme. This scheme provides a scientific basis for setting actionable emission targets at the provincial level and supports the decarbonization of the residential building sector.

The remaining sections are organized as follows: Section 2 provides a concise literature review, Section 3 outlines the methods and data sources used, Section 4 presents the results of the static and dynamic scenario analyses of emission peaks in residential buildings, Section 5 examines the optimal carbon budget allocation for each province, along with related policy recommendations, and Section 6 concludes the study with key findings.

## 2. Literature review

Studies on predicting building carbon emission peaks are primarily divided into bottom-up and top-down methods [15]. The bottom-up methods are generally categorized into statistical and engineering models. Statistical models [16, 17] mainly use regression techniques, while engineering models [18, 19] estimate energy consumption by calculating mathematical relationships based on building characteristics and energy equipment parameters. The top-down approach primarily considers macroeconomic energy-sector factors to estimate energy demand and carbon emissions [20]. Top-down approaches are typically represented by models such as the Kaya identity [21], the Impact, Population, Affluence, Technology model [22, 23], and the environmental Kuznets curve model [24]. Among these, the Kaya identity offers high universality across countries, provinces, and sectors [25, 26]. Several studies have enhanced top-down models with scenario analysis and uncertainty quantification. For example, combining the Impact, Population, Affluence, Technology models with scenario analysis [27, 28], index decomposition [29], and Monte Carlo simulations [30, 31] enables a more comprehensive assessment of future emissions. While these approaches have provided systematic insights, they have largely been applied at the national scale and often overlook spatial heterogeneity, dynamic uncertainties, and sector-specific characteristics of buildings.

At the subnational scale, studies have begun to examine provincial variations in building emissions, reflecting differences in population structure, urbanization, and energy mix. However, most have been conducted within a static framework, relying on deterministic models with limited capacity to capture the stochastic characteristics of emissions [32, 33]. In addition, few studies distinguish between urban and rural residential buildings, constraining the understanding of sectoral heterogeneity. Research focusing specifically on provincial residential emissions often lacks explicit connections to practical carbon budget or allocation decisions, thereby reducing its policy relevance [34]. These gaps underscore the need for models that are both spatially explicit and probabilistically robust.

Studies explicitly focusing on carbon budget allocation for residential buildings remain relatively scarce. Although some research has addressed the industry and transportation sectors [35, 36], far fewer have examined how emission reduction responsibilities should be distributed across provinces or building sub-sectors to meet national decarbonization targets. Moreover, only a limited number of studies have combined top-down static models with dynamic simulation methods to assess emission peaks under uncertainty—a crucial step for reliable provincial-level planning and the development of equitable carbon allocation schemes.

From this review, three major research gaps are identified. First, existing models seldom integrate spatial granularity with stochastic methods, constraining their ability to analyze uncertainties in provincial residential emissions. Second, urban-rural differences in building energy use and emissions are often overlooked, despite their pronounced divergence. Third, systematic frameworks for allocating provincial carbon reduction responsibilities within the residential sector remain underdeveloped. To address these gaps, this study integrates the Kaya identity with Monte Carlo simulation to capture inter-provincial heterogeneity, urban-rural distinctions, and uncertainties in emission peaking, thereby offering a robust and policy-relevant framework for provincial emission planning. The contributions of this study are as follows.

- **This work integrates the static Kaya identity with the dynamic Monte Carlo simulation to estimate emission trends for residential buildings in rural and urban areas.** Specifically, an emission model based on the Kaya identity is developed to estimate emission peaks for China and each province under the two scenarios. The Monte Carlo simulation is employed to incorporate parameter uncertainty, dynamically estimating the probability of emission peaks in urban and rural residential buildings across China. This integrated framework enhances both the accuracy of emission calculations and the robustness of scenario projections, providing deeper insights to support policy design.
- **This work is pioneering in proposing an optimized provincial carbon budget assessment scheme for residential buildings, under the premise of achieving net-zero emissions by 2060.** The scheme is developed based on the principle of maximizing emission reduction potential, while taking into account the practical emissions in each province. By dynamically optimizing the peaking values and peaking years for each province, this study provides essential data support for establishing provincial-level emission reduction targets and advancing the achievement of carbon neutrality.

## 3. Methods and materials

Section 3 presents the modeling of $CO_2$ emissions released by the residential building in both urban and rural China. In Section 3.1, a detailed emission model for these building operations is introduced. Section 3.2 then describes a dynamic simulation, specifically a scenario analysis, to predict the probabilities of emission peaks for residential buildings. Finally, The data sources employed in the study are described in Section 3.3.

### *3.1. Operational carbon emission model for residential buildings*

The Kaya identity [37] was employed for computing the total carbon emissions for residential buildings in rural and urban areas, respectively. Based on the selection of the five most influential factors, the Kaya identity describing the carbon emissions for the two types of buildings was established, as shown below:

$$C_{r-urban} = P \cdot f_{r-urban} \cdot e_{r-urban} \cdot K_{r-urban} \cdot U \tag{1}$$

$$C_{r-rural} = P \cdot f_{r-rural} \cdot e_{r-rural} \cdot K_{r-rural} \cdot (1-U) \tag{2}$$

where $P$ represents population size, $U$ denotes the urbanization rate; $f_{r-urban}$ and $f_{r-rural}$ reflect the per capita floor area in urban and rural residential buildings across China, respectively; $e_{r-urban}$ and $e_{r-rural}$ represent the energy intensity in urban and rural residential buildings across China, respectively; and $K_{r-urban}$ and $K_{r-rural}$ denote the comprehensive carbon emission factors in urban and rural residential buildings across China, respectively. $K_{r-urban}$ and $K_{r-rural}$ can be expressed as belows:

$$K_{r-urban} = (1 - R_{r-urban-BIPG}) \cdot (K_{coal} \cdot R_{r-urban-coal} + K_{gas} \cdot R_{r-urban-gas} + K_{electricity} \cdot R_{r-urban-electricity}) \tag{3}$$

$$K_{r-rural} = (1 - R_{r-rural-BIPG}) \cdot (K_{coal} \cdot R_{r-rural-coal} + K_{gas} \cdot R_{r-rural-gas} + K_{electricity} \cdot R_{r-rural-electricity}) \tag{4}$$

Where $K_{gas}$, $K_{coal}$, and $K_{electricity}$ are the emission factors of natural gas, coal, and electricity generation, respectively; $R_{r-urban-BIPG}$, $R_{r-urban-coal}$, $R_{r-urban-gas}$, and $R_{r-urban-electricity}$ are the proportions of self-generated energy, coal usage, natural gas usage, and electricity consumption in the energy demand of urban residential buildings, respectively; and $R_{r-rural-BIPG}$, $R_{r-rural-gas}$, $R_{r-rural-coal}$, and $R_{r-rural-electricity}$ are the proportions of self-generated energy, natural gas usage, coal usage, and electricity consumption in the energy demand of rural residential buildings, respectively.

To capture the dynamic evolution of carbon emissions, two scenarios were developed: a business-as-usual scenario and a decarbonization scenario. The key distinction between them lies in assumptions about energy intensity and energy structure. In the business-as-usual scenario, energy intensity followed the baseline trajectory. In contrast, the decarbonization scenario assumed that urban residential buildings would achieve a 30% reduction in energy intensity by 2060 relative to 2020 levels, while growth in rural residential energy intensity would be substantially moderated. Regarding energy structure, the decarbonization scenario featured significantly higher electrification rates and larger shares of self-generated energy compared with the business-as-usual scenario.

*3.2. Simulation of operational carbon in residential buildings*

To comprehensively account for potential variations in future emissions from residential buildings, constructing a dynamic emission model is essential. This model incorporates uncertainties in the future parameter distributions of each variable, enabling dynamic scenario analysis. A Monte Carlo simulation [38] was utilized to transform the static model into a dynamic one, facilitating a detailed scenario analysis of carbon peaking for residential buildings.

Firstly, probability distributions governing the variability of all parameters in the carbon emission model from Section 3.1 were established. This involved assigning standard deviation values to the normal distributions to account for uncertainties, thereby quantifying the variability, as outlined below:

$$\frac{P}{(Static)} \xrightarrow{\substack{Random \\ parameter}} \frac{P \cdot (1+\omega_P)}{(Dynamic)}, \omega_P \sim N(0, {\sigma_P}^2) \tag{5}$$

$$\frac{U}{(Static)} \xrightarrow{\substack{Random \\ parameter}} \frac{U \cdot (1+\omega_U)}{(Dynamic)}, \omega_U \sim N(0, {\sigma_U}^2) \tag{6}$$

$$\frac{f_{r-urban}}{(Static)} \xrightarrow{\substack{Random \\ parameter}} \frac{f_{r-urban} \cdot (1+\omega_{f_{r-urban}})}{(Dynamic)}, \quad \omega_{f_{r-urban}} \sim N(0, {\sigma_{f_{r-urban}}}^2) \tag{7}$$

$$\frac{f_{r-rural}}{(Static)} \xrightarrow{\substack{Random \\ parameter}} \frac{f_{r-rural} \cdot (1+\omega_{f_{r-rural}})}{(Dynamic)}, \quad \omega_{f_{r-rural}} \sim N(0, {\sigma_{f_{r-rural}}}^2) \tag{8}$$

$$\frac{e_{r-urban}}{(Static)} \xrightarrow{\substack{Random \\ parameter}} \frac{e_{r-urban} \cdot (1+\omega_{e_{r-urban}})}{(Dynamic)}, \quad \omega_{e_{r-urban}} \sim N(0, {\sigma_{e_{r-urban}}}^2) \tag{9}$$

$$\frac{e_{r-rural}}{(Static)} \xrightarrow[\quad]{\substack{Random\\ parameter}} \frac{e_{r-rural} \cdot (1+\omega_{e_{r-rural}})}{(Dynamic)}, \omega_{e_{r-rural}} \sim N(0, \sigma_{e_{r-rural}}{}^2) \tag{10}$$

$$\frac{K_{r-urban}}{(Static)} \xrightarrow[\quad]{\substack{Random\\ parameter}} \frac{K_{r-urban} \cdot (1+\omega_{K_{r-urban}})}{(Dynamic)}, \omega_{K_{r-urban}} \sim N(0, \sigma_{K_{r-urban}}{}^2) \tag{11}$$

$$\frac{K_{r-rural}}{(Static)} \xrightarrow[\quad]{\substack{Random\\ parameter}} \frac{K_{r-rural} \cdot (1+\omega_{K_{r-rural}})}{(Dynamic)}, \omega_{K_{r-rural}} \sim N(0, \sigma_{K_{r-rural}}{}^2) \tag{12}$$

where the random parameter $\omega_i$ $(i = P,\ U,\ f_{r-urban},\ f_{r-rural},\ e_{r-urban},\ e_{r-rural},\ K_{r-urban},\ K_{r-rural})$ follow normal distributions $N\,(0, \sigma_i{}^2)$, $i = P,\ U,\ f_{r-urban},\ f_{r-rural},\ e_{r-urban},\ e_{r-rural},\ K_{r-urban},\ K_{r-rural}$. This assumption aligns with prior studies on building energy and emission modeling, where normal distributions were commonly employed to represent parameter variability [39, 40]. By assigning values to the standard deviations ($\sigma$) of the normal distributions associated with the aforementioned parameters, their variability was quantified. Through repeated sampling of perturbations around the mean values, the model captured fluctuations that reflect more realistic dynamics in provincial-level emission pathways [39]. Accounting for these uncertainties, the contribution of future national population size ($P$) to the emission parameters in the residential building carbon emission model for 2021-2060 is expressed as follows:

$$\frac{P|_T}{(Dynamic)} = \frac{P|_T}{(Static)} \cdot \left(1 + \omega_P \cdot \frac{T - 2020}{2060 - 2020}\right), \omega_P \sim N(0, \sigma_P{}^2) \tag{13}$$

After quantitatively assessing the uncertainties associated with each parameter in the emission model, these uncertainties were integrated into the emissions for residential buildings. This process transformed the static carbon emission model into a dynamic one, reflecting potential variations in the emission parameters, as shown below:

$$\frac{C}{(Static)} \xrightarrow[\quad]{\substack{Random\\ parameters}} \frac{C \cdot (1+\omega)}{(Dynamic)}, \omega \sim N(0, \sigma_C{}^2) \tag{14}$$

Subsequently, 100 thousand Monte Carlo simulations were conducted to strengthen the precision and reliability of the final carbon emission peaking results. Monte Carlo simulation is a method where the results increasingly approximate the true distribution as the number of simulations grows. By performing thousands of simulations on each parameter, the statistical results were used to determine the range of variation for future emission peaks in residential buildings. Additionally, the peaking time and value for different building types were analyzed probabilistically.

The allocation framework consists of three key steps. First, provincial residential building emissions were projected under a business-as-usual scenario to establish a reference pathway. Second, a carbon-neutrality-oriented decarbonization scenario was backcast from the 2060 net-zero target, ensuring consistency with national commitments while incorporating urban-rural heterogeneity. Third, emission reductions were allocated by identifying the maximum gap between the two scenarios, thereby capturing each province's reduction potential. A detailed explanation of the allocation method is provided in Appendix C.

### *3.3. Datasets*

The core parameters employed in this study are summarized in Table 1, which provides a systematic overview of their definitions, units, and data sources. The dataset primarily draws on peer-reviewed research to ensure consistency and reliability. Specifically, population size data were taken from You et al [40]. Parameters related to per capita floor area in both urban and rural residential buildings, energy intensity values, and the urbanization rate were sourced from Zou et al [41]. Energy structure parameters—including the shares of coal, natural gas, electricity, and self-generated energy in both urban and rural contexts—as well as emission factors of coal, natural gas, and electricity generation, were obtained from Yan et al [42]. These datasets form the fundamental inputs for the modeling framework and ensure that the analysis is based on transparent and traceable parameter sources. Consistent with prior studies, this work covered thirty provinces of China, excluding Taiwan, Hong Kong, Tibet, and Macau due to data unavailability. The specific values of the main parameters are provided in Appendix D.

**Table 1.** Variable specification: definitions, units, and data collection.

| Variable | Definition | Unit | Data collection | |
|---|---|---|---|---|
| | | | Data source type | References |
| $P$ | Population size | Million persons | Research paper | Population size data: You et al [40]. |
| $f_{r-urban}$ | Per capita floor area in urban residential buildings | Square meters ($m^2$)/person | Research paper | Per capita floor area data: Zou et al [41]. |
| $f_{r-rural}$ | Per capita floor area in rural residential buildings | $m^2$/person | Research paper | Per capita floor area data: Zou et al [41]. |

| Variable | Definition | Unit | Data collection | |
|---|---|---|---|---|
| | | | Data source type | References |
| $e_{r-urban}$ | Energy intensity in urban residential buildings | kilogram of coal equivalent (kgce)/m² | Research paper | Energy intensity data: Zou et al [41]. |
| $e_{r-rural}$ | Energy intensity in rural residential buildings | kgce/m² | Research paper | Energy intensity data: Zou et al [41]. |
| $R_{r-urban-BIPG}$ | Share of self-generated energy in urban residential energy demand | % (share of total energy use) | Research paper | Proportion data: Yan et al [42]. |
| $R_{r-urban-coal}$ | Proportion of coal usage in the energy demand of urban residential buildings | % | Research paper | Proportion data: Yan et al [42]. |
| $R_{r-urban-gas}$ | Proportion of natural gas usage in the energy demand of urban residential buildings | % | Research paper | Proportion data: Yan et al [42]. |
| $R_{r-urban-electricity}$ | Proportion of electricity consumption in the energy demand of urban residential buildings | % | Research paper | Proportion data: Yan et al [42]. |
| $R_{r-rural-BIPG}$ | Share of self-generated energy in rural residential energy demand | % | Research paper | Proportion data: Yan et al [42]. |
| $R_{r-rural-coal}$ | Proportion of coal usage in the energy demand of rural residential buildings | % | Research paper | Proportion data: Yan et al [42]. |
| $R_{r-rural-gas}$ | Proportion of natural gas usage in the energy demand of rural residential buildings | % | Research paper | Proportion data: Yan et al [42]. |
| $R_{r-rural-electricity}$ | Proportion of electricity consumption in the energy demand of rural residential buildings | % | Research paper | Proportion data: Yan et al [42]. |
| $K_{coal}$ | Emission factor of coal | kilogram of carbon dioxide ($kgCO_2$)/ kgce | Research paper | Emission factor data: Yan et al [42]. |
| $K_{gas}$ | Emission factor of natural gas | $kgCO_2$/ kgce | Research paper | Emission factor data: Yan et al [42]. |
| $K_{electricity}$ | Emission factor of electricity generation | $kgCO_2$/ kgce | Research paper | Emission factor data: Yan et al [42]. |
| $U$ | Urbanization rate | % | Research paper | Urbanization rate data: Zou et al [41]. |

## 4. Results

### *4.1. Static scenario analysis of carbon emission peaks for residential buildings*

Fig. 1 illustrates the emission patterns of residential buildings from 2021 to 2060 under a static scenario for two building types: urban and rural residential buildings. Both types follow an inverted U-shaped emission curve across the two scenarios. In the business-as-usual scenario (Fig. 1 a), urban residential buildings were expected to peak in 2031 at 990 mega-tons of $CO_2$ ($MtCO_2$), while rural residential buildings (Fig. 1 b) were forecasted to peak at 450 $MtCO_2$ in 2026. In the decarbonization scenario (Fig. 1 a), urban residential buildings peaked in 2023 at 900 $MtCO_2$, and rural residential buildings (Fig. 1 b) peaked at 420 $MtCO_2$ in 2021. Consequently, under the decarbonization scenario, both building types were projected to peak in 2025.

Achieving the parameter settings of the decarbonization scenario will support the attainment of the 2060 carbon neutrality target. The decarbonization scenario presents a more ambitious potential for building decarbonization in comparison with the business-as-usual scenario. It assumes higher energy efficiency levels in buildings and a greater proportion of renewable energy in the electricity mix. Additionally, this study defines the variation range for each parameter in the emission model using a normal distribution with a ± 1-3 standard deviation change. These variations form error bands at different probability levels in the dynamic scenario simulation, representing the potential emission range for urban and rural residential buildings under varying probabilities.

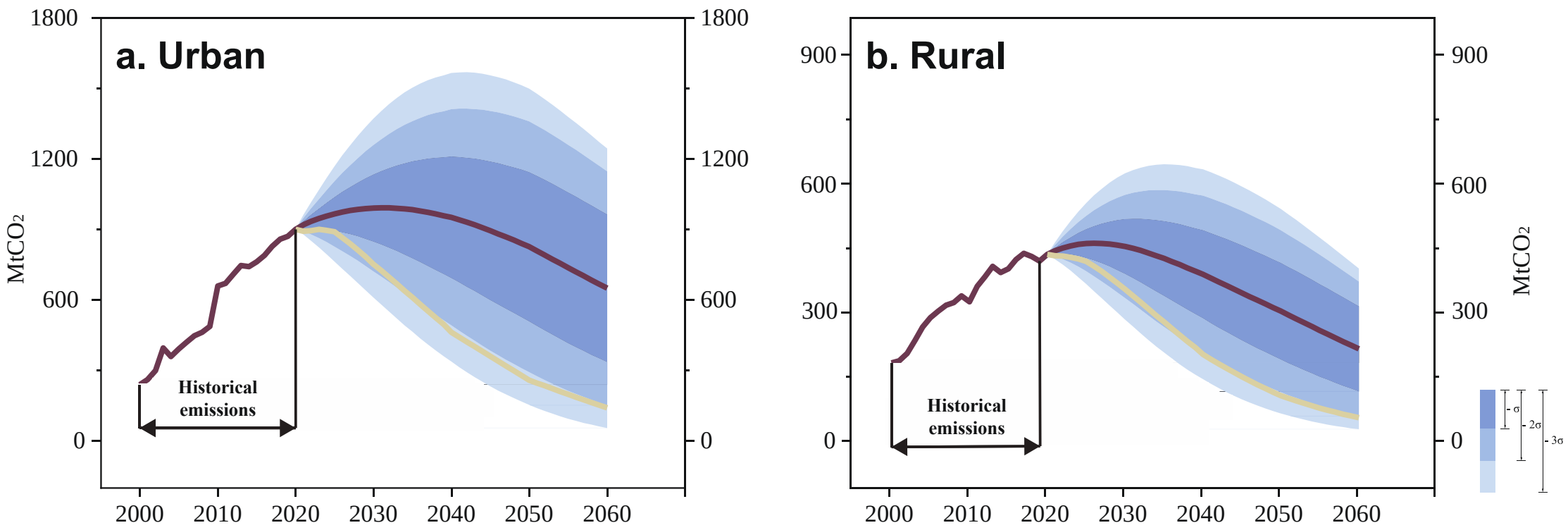

**Fig. 1.** Static carbon emission scenarios for residential building operations and prospective dynamic scenario ranges with varying probabilities in (a) urban and (b) rural China, 2000-2060.

Fig. 2 shows the emission peaks for different building types across 30 provinces in China under the business-as-usual scenario. There was significant variation in the emission peaks for urban residential buildings among the provinces, with peaking years ranging from 2021 to 2034. Regionally, in North China, Tianjin peaked first in 2021, while Shanxi was expected to peak last in 2033. In Northeast China, Liaoning was projected to peak in 2030, with other provinces peaking in 2025. In East China, Shanghai peaked in 2021, while Jiangxi was expected to reach its peak last, in 2040. Within Central China, Hubei was anticipated to peak first in 2031, with Henan peaking last in 2043. Within South China, Guangdong was anticipated to reach its peak in 2030, while Guangxi was projected to peak last in 2034. Within Southwest China, both Chongqing and Yunnan were anticipated to reach their peaks in 2033, with Guizhou peaking last in 2047. Within Northwest China, Shaanxi was anticipated to reach its peak in 2033, while Xinjiang was projected to peak last in 2050. Provincially, Shandong was expected to have the highest peak at 110 $MtCO_2$, while Hainan was expected to have the lowest at 2.84 $MtCO_2$.

The future emission peaks for rural residential buildings across the provinces also exhibited temporal differences, with the earliest peak in 2021 and the latest in 2035. In North China, Shanxi and Beijing peaked earliest in 2022, while Hebei was anticipated to reach its peak last in 2027. Within Northeast China, Heilongjiang peaked in 2022, while Liaoning was anticipated to reach its peak in 2025. Within East China, Shanghai and Jiangsu peaked in 2021, while Anhui and Jiangxi were anticipated to peak last in 2030. Within Central China, except for Hunan (2026), the remaining provinces were anticipated to reach their peaks in 2027. Within South China, Guangdong, Guangxi, and Hainan were expected to peak starting in 2029. In Southwest China, Sichuan peaked in 2024, while Yunnan was anticipated to peak last in 2027. Within Northwest China, except for Shaanxi (2026) and Xinjiang (2035), the remaining provinces were anticipated to reach their peaks in 2027. Provincially, Hebei was expected to have the highest peak at 57.16 $MtCO_2$, while Shanghai was expected to have the lowest at 1.15 $MtCO_2$.

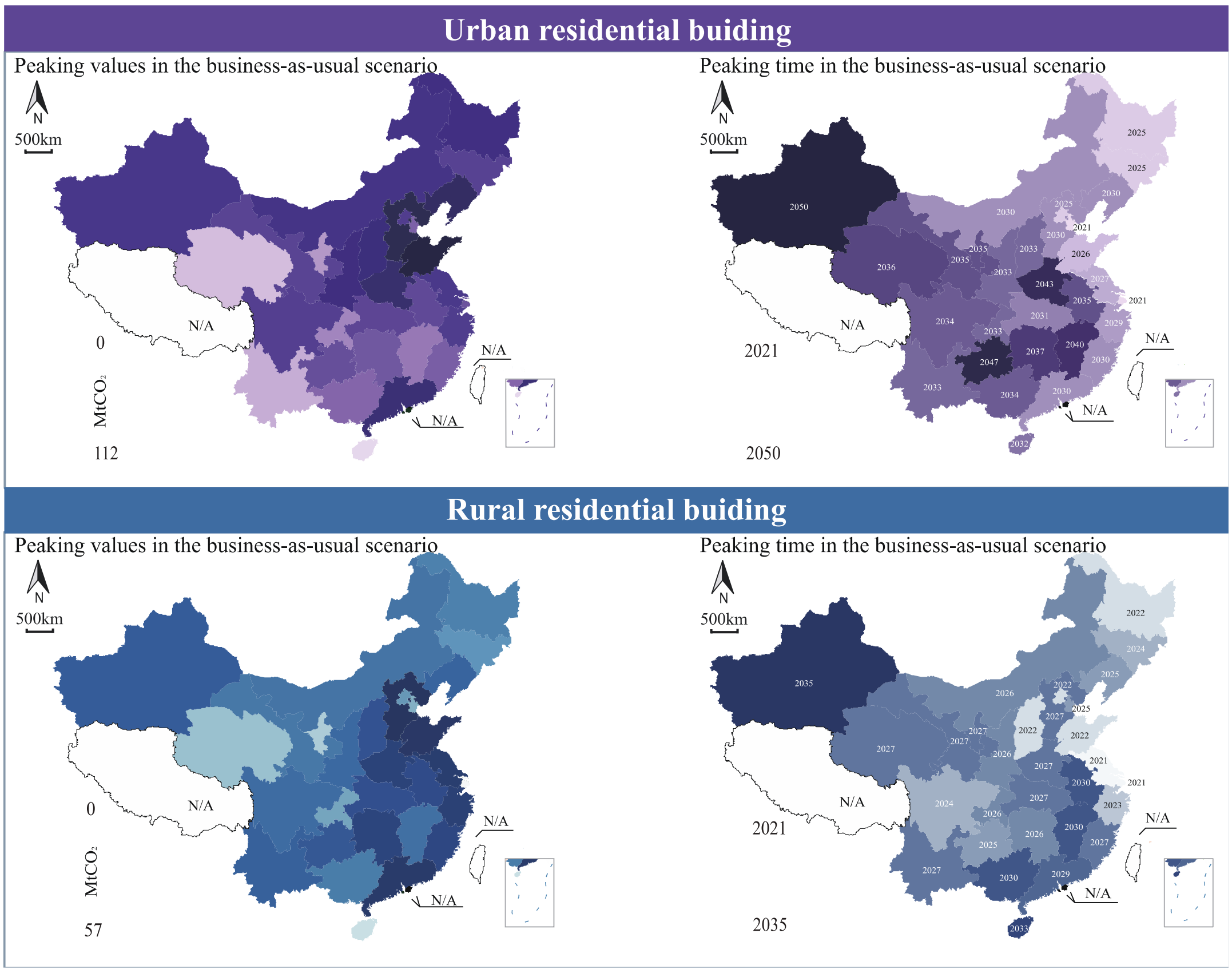

**Fig. 2.** Carbon peak values and peaking years of urban and rural residential buildings at the business-as-usual level.

Fig. 3 illustrates the emission peaks for different building types across 30 provinces in China under the decarbonization scenario within a static context. The peaking year for urban residential buildings varied by province, with the earliest peak in 2021 and the latest in 2025. Regionally, in North China, all provinces except Hebei and Shanxi peaked in 2021. In Northeast China, Liaoning peaked in 2022, while the other two provinces peaked in 2021. In East China, all provinces except Anhui, which peaked in 2022, peaked in 2021. In Central China, all provinces except Henan (2023) peaked in 2021. In South China, Guangdong peaked in 2021, while the other two provinces peaked in 2022. Within Southwest China, Chongqing and Sichuan peaked in 2021. In Northwest China, only Shaanxi peaked in 2021. At the provincial level, Shandong (110 $MtCO_2$), Hebei (67.97 $MtCO_2$), and Liaoning (64.54 $MtCO_2$) offered the highest urban residential building emissions among all provinces. The lowest were Hainan (2.46 $MtCO_2$), Qinghai (6.36 $MtCO_2$), and Yunnan (7.38 $MtCO_2$), with Shandong's peak emissions being 43 times greater than those of Ningxia.

Future emission peaks for rural residential buildings also showed significant variation across provinces, with the earliest peak in 2021 and the latest in 2025. In North and Northeast China, all provinces peaked in 2021. In East China, all provinces except Jiangxi peaked in 2021. In Central China, all provinces except Hunan (2022) peaked in 2021. In South China, Hainan, Guangdong, and Guangxi peaked successively from 2021. In Southwest China, all provinces peaked in 2021. In Northwest China, all provinces except Ningxia (2022) and Xinjiang (2024) peaked in 2021. At the provincial level, Hebei (52.57 $MtCO_2$), Shandong (36.94 $MtCO_2$), and Guangdong (27.51 $MtCO_2$) held the highest rural residential building emissions among all provinces. The lowest were Shanghai (1.12 $MtCO_2$), Hainan (1.54 $MtCO_2$), and Ningxia (1.77 $MtCO_2$), with Shandong's peak emissions being 47 times greater than those of Ningxia.

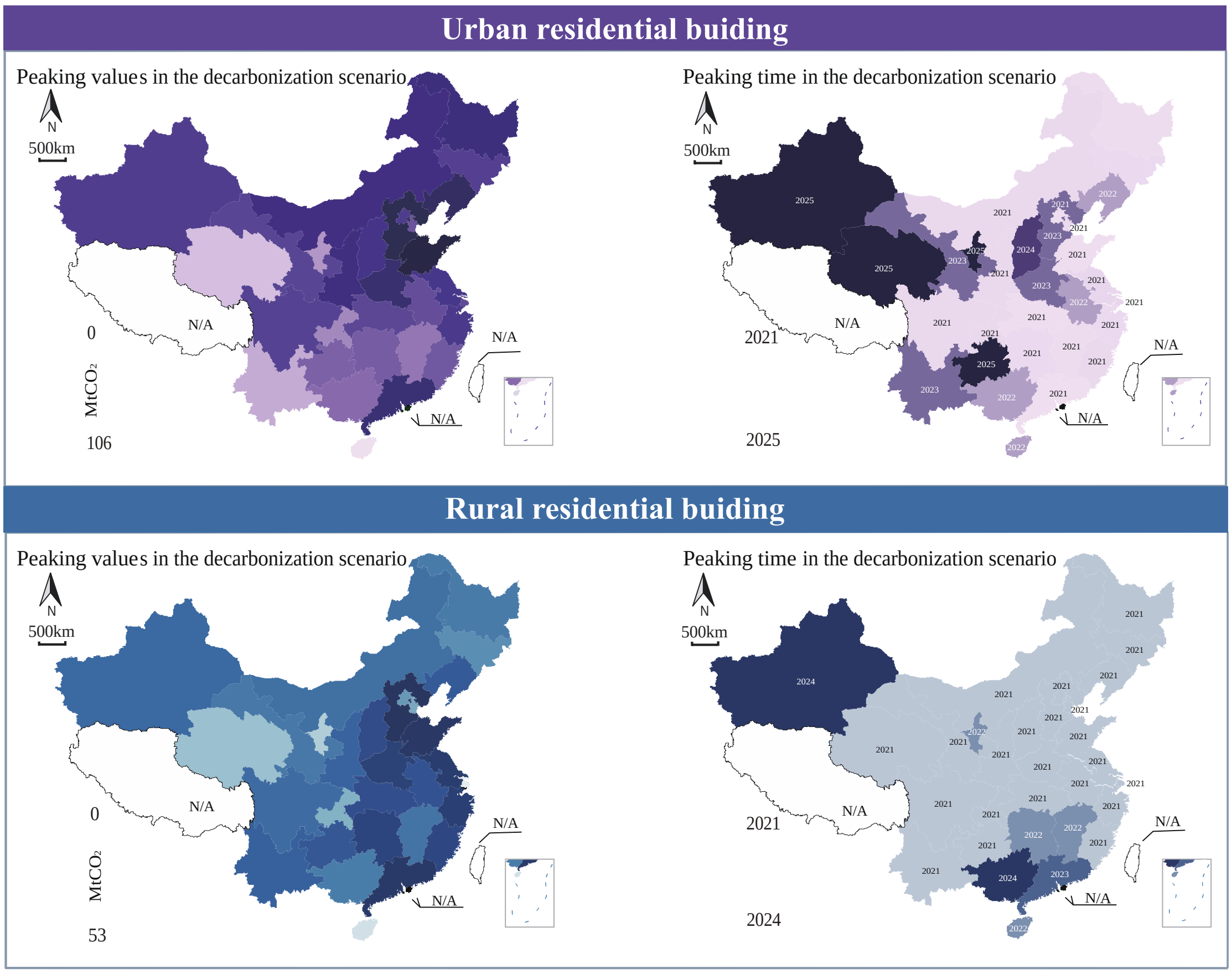

**Fig. 3.** Carbon peak values and peaking years of urban and rural residential buildings at the decarbonization level.

The significant disparities in carbon peak values and peaking years for urban and rural residential buildings across 30 provinces can largely be attributed to uneven progress in urbanization, economic development, and the adoption of energy efficiency and decarbonization technologies. This uneven development led to varying growth rates in both total and per capita building areas, despite overall expansion. Consequently, these differences gave rise to distinct emission trajectories for residential buildings across provinces. Overall, the analysis presented above models future carbon emission trends and illustrates the carbon peak status for residential buildings across different scenarios, providing a preliminary response to Question 1, as well as addressing Question 2 raised in Section 1.

### *4.2. Dynamic scenario analysis of carbon emission peaks for residential buildings*

As shown in Fig. 4, the peaks for residential building energy consumption in China were derived from 100 thousand Monte Carlo simulations of the emission model under the business-as-usual scenario. Fig. 4 a indicates that considering a 95% confidence interval of uncertainty, the energy consumption peak for urban residential buildings was estimated at 520 (±60) mega-tons of coal equivalent (Mtce), while for rural China, it was 260 (±10) Mtce (see Fig. 4 b). Thus, the energy consumption of urban residential buildings was approximately twice that of rural residential buildings. Additionally, the peaking years for urban and rural China differed. Energy consumption for urban residential buildings was projected to peak in 2043 (±7.6 years) (Fig. 4 a), whereas for rural residential buildings, the peak was estimated for 2029 (±5.0 years) (Fig. 4 b). Consequently, urban residential buildings were identified as the primary driver in the early peaking of residential building energy consumption.

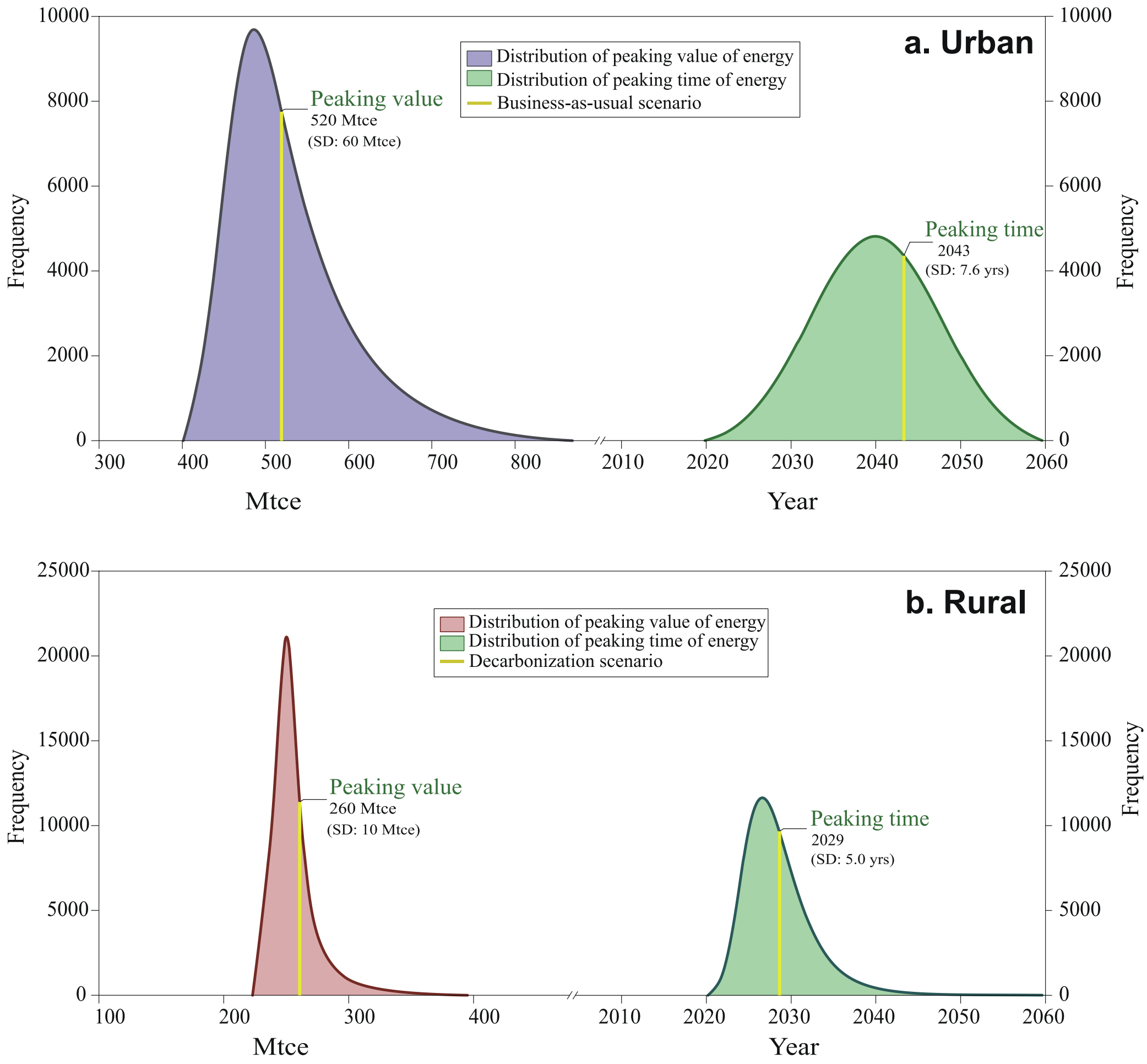

**Fig. 4.** Distribution of peak values and peaking years of energy consumption for residential buildings in (a) urban China and (b) rural China.

As shown in Fig. 5, the emission peaking years and values for residential buildings under the business-as-usual scenario were derived from 100 thousand Monte Carlo simulations. Fig. 5 a indicates that, considering uncertainty, the peaking value for urban residential buildings was estimated at 990 (±70) $MtCO_2$, while for rural residential buildings, it was 450 (±0.2) $MtCO_2$ (see Fig. 5 b). The emissions from urban and rural residential buildings showed a similar ratio to their energy consumption, approximately 2:1, with urban buildings accounting for the majority of $CO_2$ emissions. Furthermore, there were differences in the peaking years. The projected peaking year for urban residential buildings was 2031 (±4.7 years) (Fig. 5 a), while rural residential buildings were estimated to peak in 2026 (±2.6 years) (Fig. 5 b). Thus, rural residential buildings were expected to reach carbon peaking before 2030, while the challenge for urban buildings to do so by 2030 appears more manageable. To ensure the robustness of the dynamic scenario analysis, a comprehensive sensitivity analysis was performed at the national level to examine how critical parameters influence both the level and timing of the residential building carbon emissions peak. Further details can be found in Appendix E.

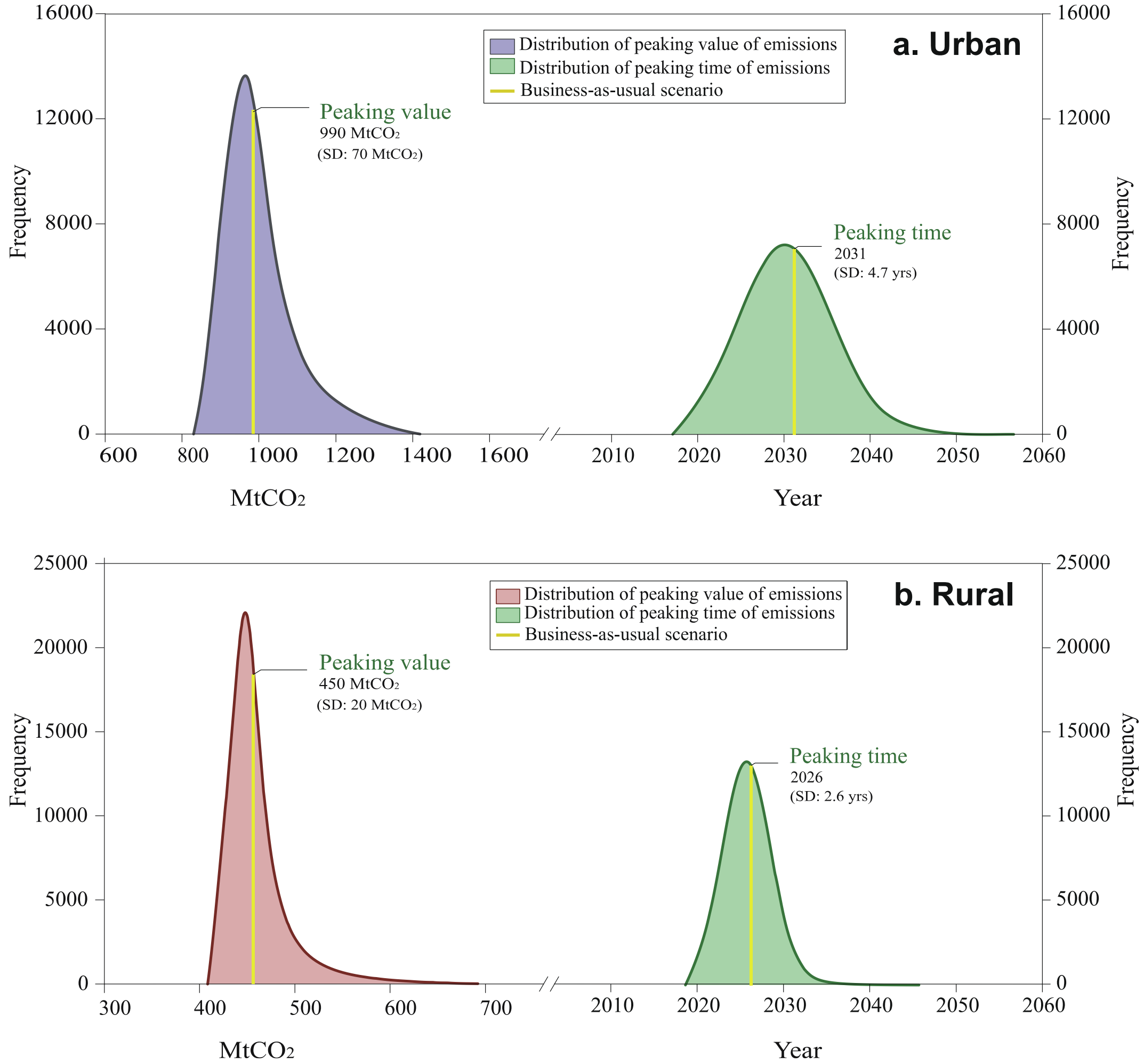

**Fig. 5.** Distribution of peak levels and peaking years of operational carbon for residential buildings in (a) urban China and (b) rural China.

## 5. Discussion

### *5.1. Probabilities of residential buildings toward a decarbonization pathway*

Fig. 6 presents the probabilistic analysis results of peak carbon emissions under decarbonization scenarios for two building types in China. For urban residential buildings, the probability of reaching the peak in 2023 under the decarbonization scenario was 3.386%, with a 0.021% chance of achieving the peaking values interval. The corresponding peak emissions were estimated at 900 $MtCO_2$, a reduction of 90 $MtCO_2$ in comparison with the business-as-usual scenario. Similarly, for rural residential buildings, the probability of reaching the peak in 2021 under the decarbonization scenario was 0.954%, with a 0.003% chance of achieving the peaking values interval. The corresponding peak emissions were estimated at 420 $MtCO_2$, indicating a reduction of 30 $MtCO_2$ in comparison with the business-as-usual scenario. Overall, the analysis presented in Sections 4.1, 4.2, and 5.1, illustrating operational carbon trends, particularly carbon peaks, in residential buildings nationwide, fully addresses Question 1 raised in Section 1.

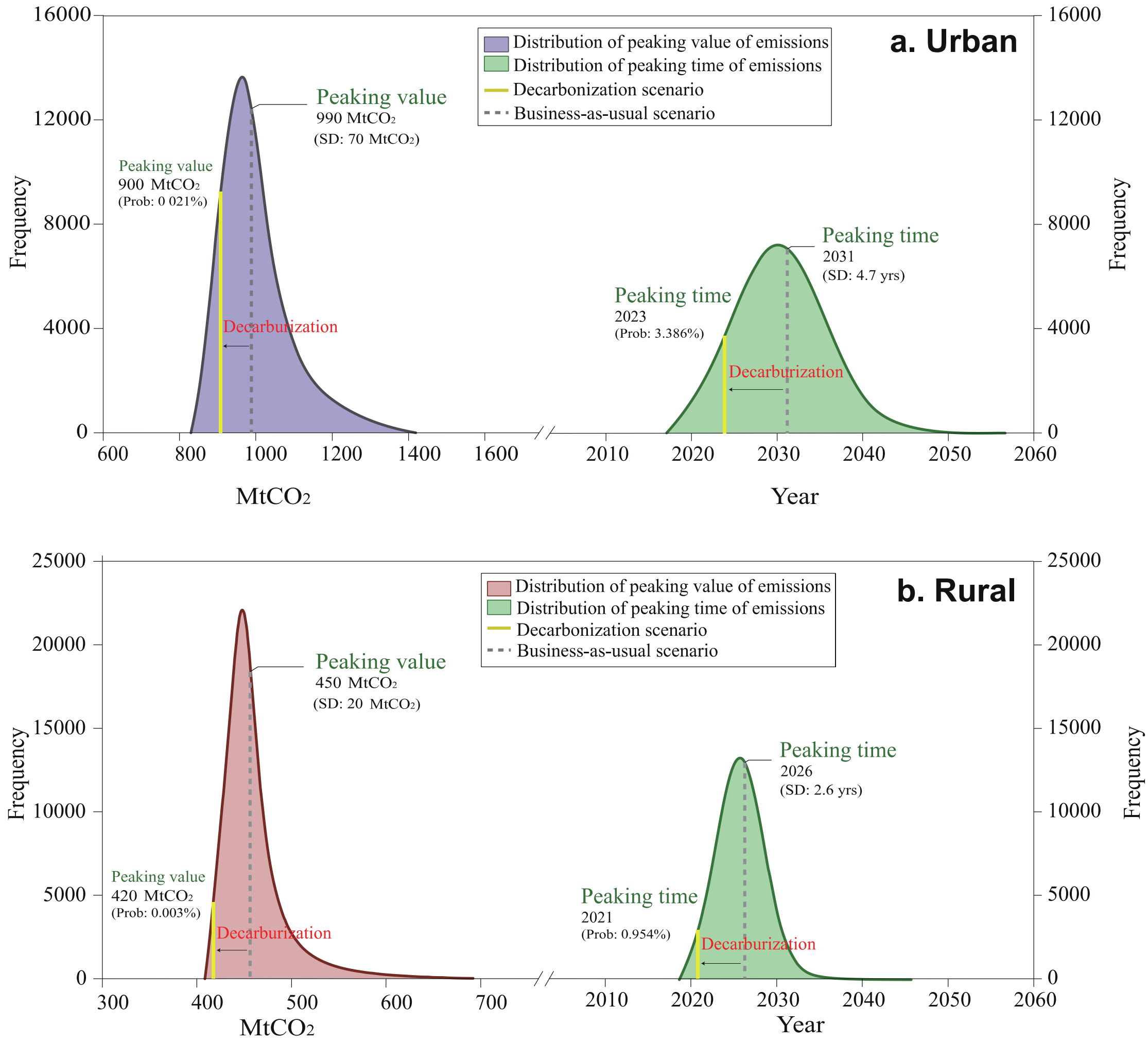

**Fig. 6.** Probabilities of residential buildings' carbon peak transitioning from the business-as-usual scenario to the decarbonization scenario in (a) urban China and (b) rural China.

### *5.2. Provincial carbon budget schemes to advance net-zero ambitions in residential buildings*

Figs. 7 and 8 present the probabilistic analysis of the peaking values and peaking years for urban residential buildings within 30 provinces in China. Future urban residential building emissions were modeled using Monte Carlo simulations under the business-as-usual scenario, determining a 0.021% probability of reaching peaking value and time interval under the decarbonization scenario. Using this probability, additional Monte Carlo simulations were performed to assess urban residential building emissions across 30 provinces, within the constraints of achieving the nationwide decarbonization goals. This analysis produced the peaking values of future emissions for urban residential buildings, which aligned with China's decarbonization objectives.

As shown in Fig. 7, at the provincial scale, the greatest emission reduction demands were observed in Henan (16.74 $MtCO_2$), Xinjiang (12.59 $MtCO_2$), and Hebei (8.46 $MtCO_2$), meanwhile Hainan (0.43 $MtCO_2$), Shanghai (0.51 $MtCO_2$), and Tianjin (0.57 $MtCO_2$) had the lowest emission reduction requirements. From a regional perspective, the northwest bore the greatest emission reduction responsibility, totaling 24.76 $MtCO_2$ across five provinces. The average required reduction in Central China was the highest among regions, at 7.53 $MtCO_2$ for its three provinces, while South China had the lowest reduction demand of 8.38 $MtCO_2$, with the lowest average reduction of 2.79 $MtCO_2$ across its three provinces.

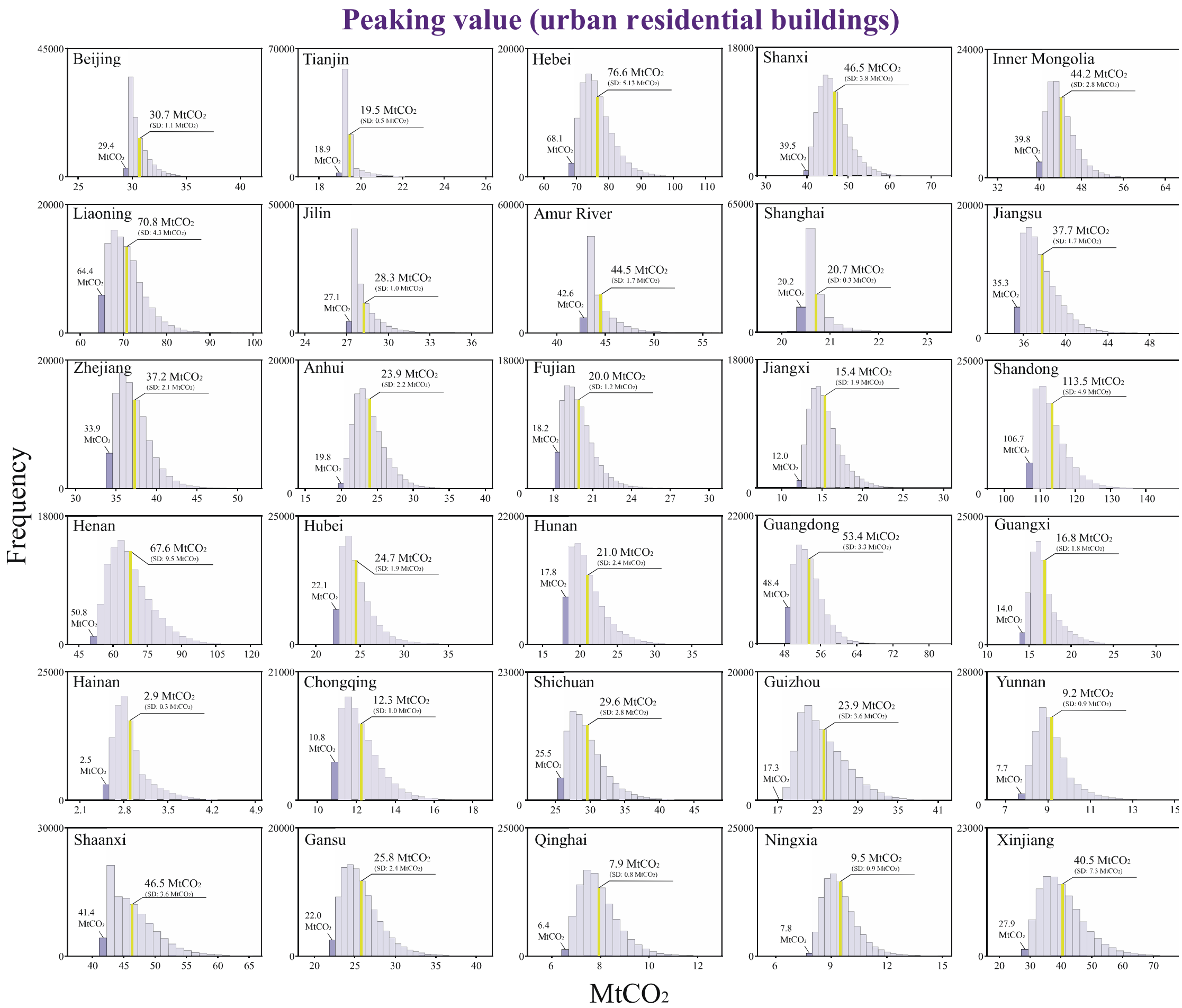

**Fig. 7.** Distribution of carbon peak values for urban residential buildings across 30 provinces.

Fig. 8 shows that achieving decarbonization goals required provinces to advance the carbon peaking time for urban residential buildings in comparison with the business-as-usual scenario. At the provincial level, the provinces needing the most substantial acceleration in peaking time were Xinjiang (by 14 years), Guizhou (by 13 years), and Hunan and Jiangxi (both by 12 years), while Shanghai required the least advancement, with just a 1-year shift.

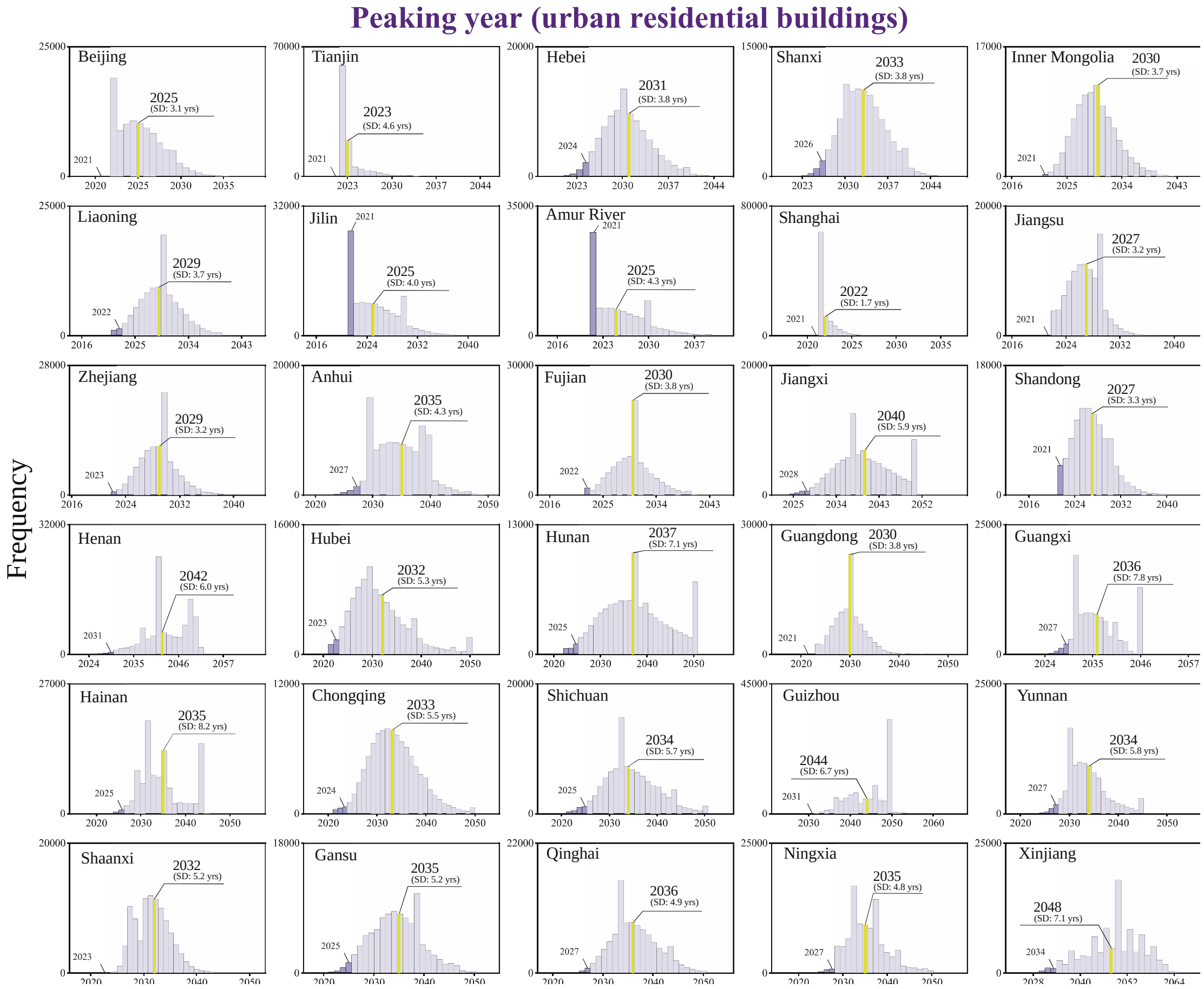

**Fig. 8.** Distribution of carbon peaking years for urban residential buildings across 30 provinces.

Figs. 9 and 10 present the probabilistic analysis of peaking values and peaking times for rural residential buildings across 30 provinces in China. From Fig. 9, among all provinces, Gansu, Hebei, and Guangdong held the highest emission reduction demands, at 9.87, 4.86, and 3.37 $MtCO_2$ respectively. Meanwhile, the provinces with the least required reductions were Shanghai (0.07 $MtCO_2$), Ningxia (0.15 $MtCO_2$), and Qinghai (0.19 $MtCO_2$). Regionally, the northwest bore the greatest emission reduction responsibility, with a total reduction requirement of 13.38 $MtCO_2$ across five provinces, and the highest average reduction responsibility at 2.68 $MtCO_2$. In contrast, the northeast region had the smallest total reduction responsibility at 1.56 $MtCO_2$, with the lowest average reduction responsibility at 0.52 $MtCO_2$.

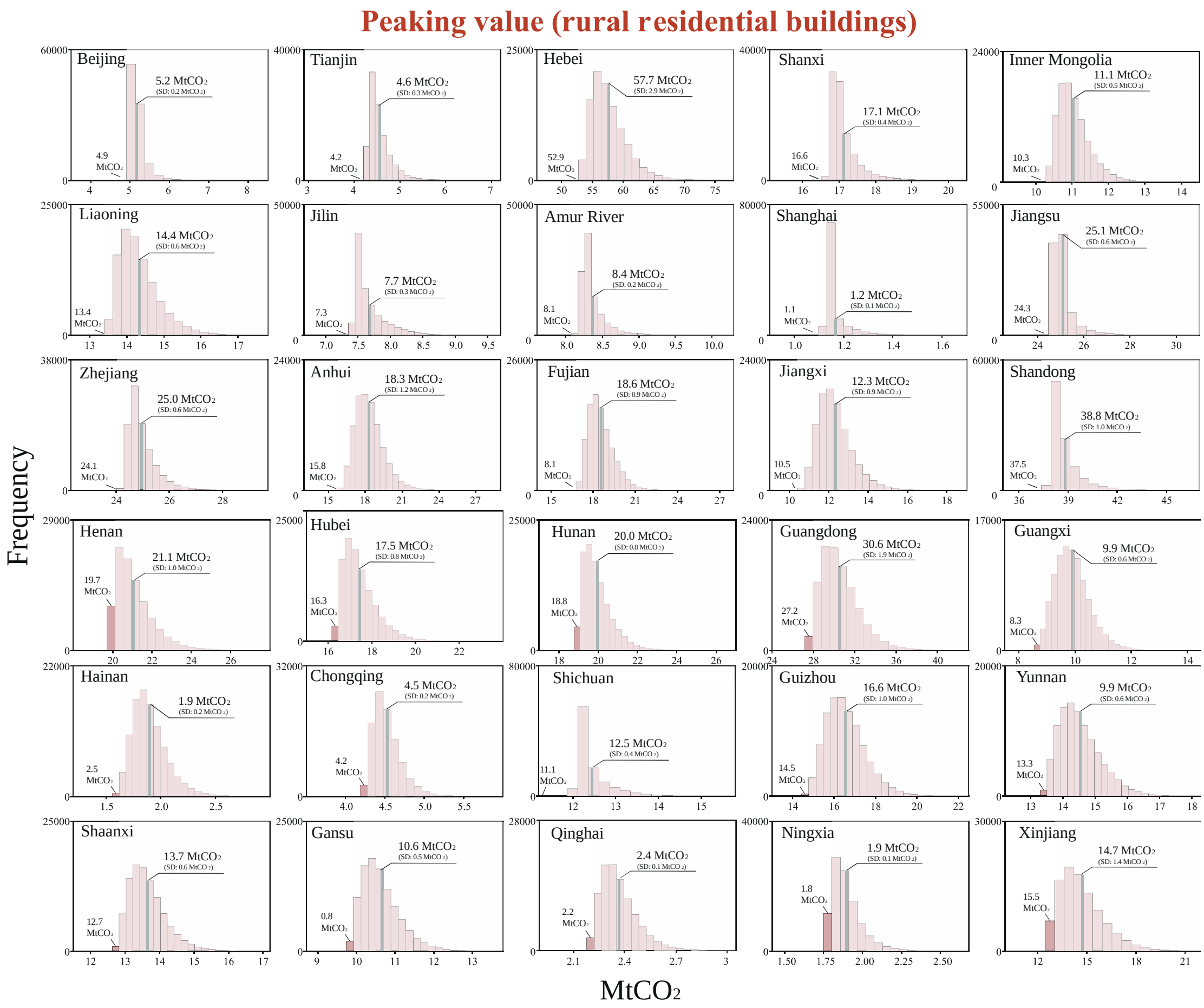

**Fig. 9.** Distribution of carbon peak values for rural residential buildings across 30 provinces.

Fig. 10 clearly demonstrates that achieving decarbonization goals required provinces to advance the carbon peaking time for rural residential buildings in comparison with the business-as-usual scenario. At the provincial level, the provinces requiring the earliest advancements in peaking time were Xinjiang (by 9 years), Hainan (by 8 years), and Jiangxi (by 7 years), while Jiangsu, Beijing, and Heilongjiang required the least advancement, with just 2-year shifts. In summary, the preceding discussion in Section 5.2 portrays the provincial carbon budget schemes and responds to Question 3 raised in Section 1.

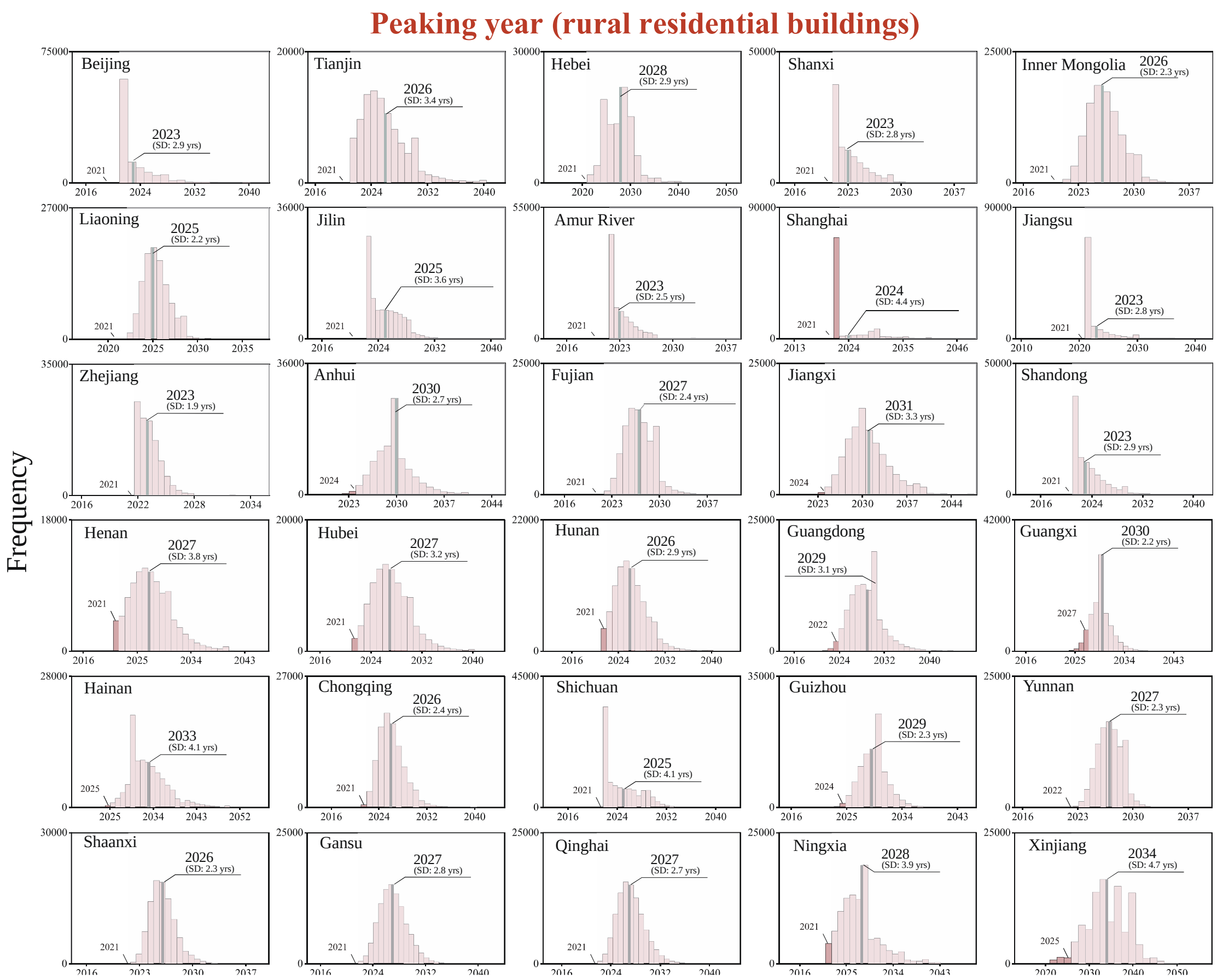

**Fig. 10.** Distribution of carbon peaking years for rural residential buildings across 30 provinces.

### *5.3. Policy recommendations*

From an international perspective, promoting energy efficiency and emission reduction for residential buildings is essential for developed countries to proactively address the climate crisis. This effort not only contributes to lowering energy consumption and greenhouse gas emissions but also aligns with sustainable development strategies [43, 44]. From a national perspective, improving energy efficiency and decarbonizing buildings is a widely accpeted strategy to address climate change. It supports the transition to a new phase of economic development, accelerates industrial optimization, and fosters new urbanization [45]. Additionally, it drives the energy revolution and curbs irrational energy use. Given the current shortcomings in nationwide and provincial planning for building energy efficiency and emission reduction, this section provides policy recommendations for China in the following areas.

a. **Addressing regional disparities and enabling an equitable transition.** Northwest China faces the highest emission reduction responsibilities, yet its relatively weaker fiscal capacity, institutional readiness, and socioeconomic conditions make large-scale transitions especially challenging. Achieving a just and feasible pathway therefore requires a comprehensive support framework that integrates both financial and technical measures: **(i)** targeted central transfers, concessional green lending, and bond guarantees to reduce the cost of capital [46]; **(ii)** inter-provincial technology cooperation tailored to cold-arid climates, including advanced building envelopes, efficient heating systems, and adaptive design standards [47]; **(iii)** systematic capacity-building for local governments and small and medium-sized enterprises to strengthen implementation and enforcement capacity [48]; and **(iv)** leveraging the region's abundant renewable energy resources to provide low-carbon electricity and accelerate heat pump adoption [49]. Together, these measures can align socioeconomic feasibility with technical viability, ensuring that emission reductions in less-developed regions contribute equitably to national decarbonization goals.

b. **Tailored provincial strategies aligned with measured potentials.** The provinces with the highest emission reduction responsibilities are Henan (16.74 $MtCO_2$), Xinjiang (12.59 $MtCO_2$), Gansu (9.87 $MtCO_2$), Hebei (8.46 $MtCO_2$), and Guangdong (3.37 $MtCO_2$), requiring differentiated pathways tailored to their respective contexts. In **Henan**, the focus should be on large-scale retrofits of existing multi-family buildings, upgrading envelope insulation, airtightness, and windows, alongside the acceleration of efficient district heating operations, digital controls, and widespread deployment of heat pumps and clean cooking technologies [50]. For **Xinjiang**, emission reductions must follow a cold-arid urban pathway that emphasizes high-performance envelopes, building-level heat recovery, advanced metering and control systems, and staged conversion from coal-based to renewable or electric heating; dedicated green-finance facilities and performance-based contracts are essential to address fiscal and institutional constraints [51]. **Hebei** requires a dual-track strategy: in urban areas, stricter enforcement of building energy codes, district system upgrades, and demand-response integration to manage peak loads; in rural areas, promotion of clean heating technologies such as air-source heat pumps or biomass-gasification, coupled with rooftop photovoltaics, storage, and standardized low-cost retrofits, supported by coordination with industrial transition programs [52]. **Gansu**'s rural sector requires a "low-cost deep-impact" package, combining prefabricated insulation kits, solar thermal water heating, efficient stoves, village-level microgrids, and targeted subsidies paired with results-based financing, while simultaneously building local capacity through installer and auditor training [53]. Finally, **Guangdong** can leverage its stronger fiscal base to pioneer smart home energy management, comprehensive retrofits, and photovoltaic generation combined with energy storage integration, supported by green mortgages, property assessed clean energy type mechanisms, and provincial product and service catalogs to accelerate diffusion [54].

c. **Phased implementation consistent with feasibility and capacity. Short term (to 2030):** Prioritize pilots in high-potential provinces, strengthen compliance with existing energy codes, and deploy "quick-win" retrofits such as efficient lighting, the heating, ventilation, and air conditioning (HVAC) controls, and basic envelope improvements. At the same time, establish provincial-level green finance facilities and standardized energy service company contracts [55]. **Medium term (2030-2045):** Scale up comprehensive retrofits, accelerate electrification of space and water heating, expand rooftop photovoltaics and community renewable projects, and reinforce distribution networks. Institutionalize results-based subsidies to ensure verified energy savings [56]. **Long term (2045-2060):** Complete the transition to net-zero-ready residential buildings through widespread adoption of digital energy management systems, high renewable penetration, and integrated storage solutions. Sustained fiscal transfers and technical support will be essential to ensure that less-developed regions achieve parity in service quality and keep pace with the national decarbonization trajectory [57].

d. **Optimizing urban planning, revitalizing existing building stock, and avoiding large-scale demolitions** are critical strategies for reducing carbon emissions in the built environment. Improved urban planning and infrastructure can enhance supporting services such as healthcare, education, and transportation while lowering vacancy rates and preventing carbon lock-in caused by poor planning and long building lifespans [58]. Upgrading existing buildings and promoting green renovation models further reduce the demand for new construction and mitigate energy use linked to the urban heat island effect [59]. At the same time, stricter controls on super high-rise projects and building heights are recommended to limit operational energy consumption from elevators and water supply systems [60]. Urban renewal should emphasize preservation and renovation over demolition, supported by robust evaluation mechanisms, clear standards, and practical implementation guidelines. Together, these measures can ensure low-carbon development while safeguarding the quality and functionality of urban spaces.

e. **Promoting intelligent control technologies and improving the energy efficiency of building equipment** are key strategies for reducing energy consumption in the building sector. This study recommends optimizing the operation and regulation of major energy-intensive systems, such as air conditioning and lighting [61]. It further advocates for the large-scale adoption of high-efficiency HVAC systems, light-emitting diode lighting, heat pumps, and other advanced energy-saving technologies. In addition, strengthening mandatory efficiency labeling standards and encouraging product designs that prioritize energy savings are essential. While high-efficiency equipment often carries higher costs due to auxiliary features [62], restricting low-efficiency appliances from entering the market could incentivize manufacturers to transition toward low-carbon product design and manufacturing.

f. **Improving subsidy mechanisms for energy-saving technologies, refining energy market pricing, and strengthening green finance** are essential to enhance the economic viability of low-carbon technologies while increasing the relative costs of carbon-intensive products and services [63]. These measures can help overcome behavioral lock-in, facilitate the adoption of energy-saving initiatives, and promote emission reductions in residential buildings. Accordingly, this study recommends assessing the potential and cost-effectiveness of low-carbon technologies, identifying key household products and services, and establishing a promotion directory focused on green buildings and renewable energy installations. A diversified subsidy framework—combining financial grants, concessional loans, and tax incentives—should be complemented by innovative green financial instruments developed by banks and insurance institutions under manageable risk conditions [64]. At the same time, the energy-saving services industry should be expanded through energy performance contracting models (e.g., guaranteed savings and shared savings) as well as one-stop integrated services covering consulting, diagnostics, planning, financing, retrofitting, and operational management.

## 6. Conclusion

This study evaluated the carbon emissions from residential building operations in China through 2060 and proposed a dynamically optimized provincial carbon budget scheme to facilitate carbon neutrality. Specifically, it developed an emission model for residential buildings from the national to the provincial level, projected emission peaks in China and each province under both the business-as-usual and decarbonization scenarios, and estimated the probabilities of emission peaks in urban and rural China using Monte Carlo simulation. Furthermore, an optimized provincial carbon budget assessment scheme was proposed to maximize emission reductions and advance the net-zero goal. The principle findings are summarized below.

### *6.1. Principal findings*

- **Under the business-as-usual scenario, national urban residential buildings were projected to peak at 990 (±0.7) $MtCO_2$ in 2031 (±4.7 years), while rural residential buildings were expected to peak at 450 (±0.2) $MtCO_2$ in 2026 (±2.6 years).** Within the business-as-usual scenario, Shandong exhibited the highest peaking value for urban residential buildings at 110 $MtCO_2$, while Hainan showed the lowest peaking value at 2.84 $MtCO_2$. For rural residential buildings, Hebei was projected to reach the highest peaking value at 57.2 $MtCO_2$, whereas Shanghai would have the lowest peaking value at 1.15 $MtCO_2$.
- **Under the decarbonization scenario, the peak value for urban residential buildings was estimated at 900 $MtCO_2$, representing a reduction of 90 $MtCO_2$ compared to the business-as-usual scenario.** The probability of urban residential buildings reaching this peak in 2023 was 3.386%, with a 0.021% probability of achieving the peak value interval. Rural residential buildings were projected to peak at 420 $MtCO_2$ in the decarbonization scenario, reflecting a reduction of 30 $MtCO_2$ relative to the business-as-usual scenario. The probability of rural residential buildings reaching their peak in 2021 was 0.954%, with a 0.003% probability of achieving the peak value interval.
- **The key contributors to emission reduction needs included Henan, Xinjiang, Gansu, Hebei, and Guangdong, with Northwest China collectively bearing the greatest** **responsibility.** For urban residential buildings, the provinces with the highest emission reduction potential were Henan (16.74 $MtCO_2$), Xinjiang (12.59 $MtCO_2$), and Hebei (8.46 $MtCO_2$), while those with the lowest potential were Hainan (0.43 $MtCO_2$), Shanghai (0.51 $MtCO_2$), and Tianjin (0.57 $MtCO_2$). For rural residential buildings, the provinces requiring the most significant emission reductions were Gansu (9.87 $MtCO_2$), Hebei (4.86 $MtCO_2$), and Guangdong (3.37 $MtCO_2$), while the lowest demands were observed in Shanghai (0.07 $MtCO_2$), Ningxia (0.15 $MtCO_2$), and Qinghai (0.19 $MtCO_2$).

### *6.2. Upcoming works*

Several limitations of this study warrant further investigation. This study focused solely on estimating operational-phase energy use and associated emissions of residential buildings. However, considering the full life cycle emissions of residential buildings would provide a more comprehensive assessment. Additionally, the primary aim of predicting emission trends for residential buildings in urban and rural China is to support the net-zero target. Therefore, assessing the decarbonization trends for other building types, such as commercial buildings, is also crucial. Furthermore, as buildings progressively decarbonize, incorporating a cost-benefit analysis when implementing decarbonization measures becomes essential.

## Declarations

### Competing interests

None.

### Funding

First author appreciates the National Planning Office for Philosophy and Social Science Foundation (24BJY129).

### Authors' contributions

Conceptualization, M.M.; Methodology, M.M., H.Y., and Z.M.; Software, M.M., and H.Y.; Validation, Z.M., and N.Z.; Writing-Original Draft, M.M., and H.Y.; Writing-Review & Editing, M.M., Z.M., N.Z., C.Z., and H.Y.; Funding Acquisition, M.M., and N.Z.; Supervision, M.M., N.Z., and C.Z..

### Acknowledgements

None.

## Appendix

The appendix is available in the supplementary material (e-component) accompanying this submission.

## References

[1] Lamb N, Elmes D. Increasing heat pump adoption: analysing multiple perspectives on preparing homes for heat pumps in the UK. Carbon Neutrality 2024;3:10.

[2] Ma M, Zhou N, Feng W, Yan J. Challenges and opportunities in the global net-zero building sector. Cell Reports Sustainability 2024;1:100154.

[3] Camarasa C, Mata É, Navarro JPJ, Reyna J, Bezerra P, Angelkorte GB, et al. A global comparison of building decarbonization scenarios by 2050 towards 1.5–2 C targets. Nature Communications 2022;13:3077.

[4] Zhang S, Ma M, Zhou N, Yan J, Feng W, Yan R, et al. Estimation of global building stocks by 2070: Unlocking renovation potential. Nexus 2024;1:100019.

[5] Wang Z, Sun Y, Kong H, Xia-Bauer C. An in-depth review of key technologies and pathways to carbon neutrality: classification and assessment of decarbonization technologies. Carbon Neutrality 2025;4:15.

[6] Zheng H, Song M, Shen Z. The evolution of renewable energy and its impact on carbon reduction in China. Energy 2021;237:121639.

[7] Berrill P, Wilson EJ, Reyna JL, Fontanini AD, Hertwich EG. Decarbonization pathways for the residential sector in the United States. Nature Climate Change 2022;12:712-718.

[8] Xu W, Sun D, Dong J, Huang S, Yang S. Technical pathways of dual carbon control in China's buildings sector. Carbon Neutrality 2022;1:18.

[9] Jing R, Wang X, Zhao Y, Zhou Y, Wu J, Lin J. Planning urban energy systems adapting to extreme weather. Advances in Applied Energy 2021;3:100053.

[10] González-Torres M, Pérez-Lombard L, Coronel JF, Maestre IR, Paolo B. Activity and efficiency trends for the residential sector across countries. Energy and Buildings 2022;273:112428.

[11] Fang K, Tang Y, Zhang Q, Song J, Wen Q, Sun H, et al. Will China peak its energy-related carbon emissions by 2030? Lessons from 30 Chinese provinces. Applied Energy 2019;255:113852.

[12] Röck M, Saade MRM, Balouktsi M, Rasmussen FN, Birgisdottir H, Frischknecht R, et al. Embodied GHG emissions of buildings – The hidden challenge for effective climate change mitigation. Applied Energy 2020;258:114107.

[13] Solarin SA. Convergence in CO 2 emissions, carbon footprint and ecological footprint: evidence from OECD countries. Environmental Science and Pollution Research 2019;26:6167-6181.

[14] Yue X, Deane JP, O'Gallachoir B, Rogan F. Identifying decarbonisation opportunities using marginal abatement cost curves and energy system scenario ensembles. Applied Energy 2020;276:115456.

[15] Yan R, Zhou N, Ma M, Mao C. India's residential space cooling transition: Decarbonization ambitions since the turn of millennium. Applied Energy 2025;391:125929.

[16] Zhou N, Price L, Yande D, Creyts J, Khanna N, Fridley D, et al. A roadmap for China to peak carbon dioxide emissions and achieve a 20% share of non-fossil fuels in primary energy by 2030. Applied Energy 2019;239:793-819.

[17] Kaltenegger O. What drives total real unit energy costs globally? A novel LMDI decomposition approach. Applied Energy 2020;261:114340.

[18] Zeynali S, Rostami N, Feyzi M, Mohammadi-Ivatloo B. Multi-objective optimal planning of wind distributed generation considering uncertainty and different penetration level of plug-in electric vehicles. Sustainable Cities and Society 2020;62:102401.

[19] Li R, Li L, Wang Q. The impact of energy efficiency on carbon emissions: evidence from the transportation sector in Chinese 30 provinces. Sustainable Cities and Society 2022;82:103880.

[20] Vaisi S, Firouzi M, Varmazyari P. Energy benchmarking for secondary school buildings, applying the Top-Down approach. Energy and Buildings 2023;279:112689.

[21] Lin Y, Ma L, Li Z, Ni W. The carbon reduction potential by improving technical efficiency from energy sources to final services in China: An extended Kaya identity analysis. Energy 2023;263:125963.

[22] Eibinger T, Deixelberger B, Manner H. Panel data in environmental economics: Econometric issues and applications to IPAT models. Journal of Environmental Economics and Management 2024;125:102941.

[23] Tan J, Peng S, Liu E. Spatio-temporal distribution and peak prediction of energy consumption and carbon emissions of residential buildings in China. Applied Energy 2024;376:124330.

[24] Makarov I, Alataş S. Production- and consumption-based emissions in carbon exporters and importers: A large panel data analysis for the EKC hypothesis. Applied Energy 2024;363:123063.

[25] Yuan H, Ma X, Ma M, Ma J. Hybrid framework combining grey system model with Gaussian process and STL for CO2 emissions forecasting in developed countries. Applied Energy 2024;360:122824.

[26] Fang K, Li C, Tang Y, He J, Song J. China's pathways to peak carbon emissions: New insights from various industrial sectors. Applied Energy 2022;306:118039.

[27] Catrini P, Curto D, Franzitta V, Cardona F. Improving energy efficiency of commercial buildings by Combined Heat Cooling and Power plants. Sustainable Cities and Society 2020;60:102157.

[28] Dong K, Dong X, Dong C. Determinants of the global and regional CO2 emissions: what causes what and where? Applied Economics 2019;51:5031-5044.

[29] Lin B, Raza MY. Analysis of electricity consumption in Pakistan using index decomposition and decoupling approach. Energy 2021;214:118888.

[30] Li Y, Wang J, Deng B, Liu B, Zhang L, Zhao P. Emission reduction analysis of China's building operations from provincial perspective: factor decomposition and peak prediction. Energy and Buildings 2023;296:113366.

[31] Robati M, Oldfield P. The embodied carbon of mass timber and concrete buildings in Australia: An uncertainty analysis. Building and Environment 2022;214:108944.

[32] Deng Y, Ma M, Zhou N, Ma Z, Yan R, Ma X. China's plug-in hybrid electric vehicle transition: An operational carbon perspective. Energy Conversion and Management 2024;320:119011.

[33] Taniguchi-Matsuoka A, Shimoda Y, Sugiyama M, Kurokawa Y, Matoba H, Yamasaki T, et al. Evaluating Japan's national greenhouse gas reduction policy using a bottom-up residential end-use energy simulation model. Applied Energy 2020;279:115792.

[34] Yan Y, Zhang H, Long Y, Zhou X, Liao Q, Xu N, Liang Y. A factor-based bottom-up approach for the long-term electricity consumption estimation in the Japanese residential sector. Journal of Environmental Management 2020;270:110750.

[35] Liu H, Chen Y, Wu J, Pan Y, Song Y. Allocation of CO2 emission quotas for industrial production in Industry 4.0: Efficiency and equity. Computers & Industrial Engineering 2024;194:110375.

[36] Heinold A, Meisel F. Emission limits and emission allocation schemes in intermodal freight transportation. Transportation Research Part E: Logistics and Transportation Review 2020;141:101963.

[37] Gudipudi R, Rybski D, Lüdeke MKB, Zhou B, Liu Z, Kropp JP. The efficient, the intensive, and the productive: Insights from urban Kaya scaling. Applied Energy 2019;236:155-162.

[38] Liu C, Zheng X, Yang H, Tang W, Sang G, Cui H. Techno-economic evaluation of energy storage systems for concentrated solar power plants using the Monte Carlo method. Applied Energy 2023;352:121983.

[39] Wang H, Lu X, Deng Y, Sun Y, Nielsen CP, Liu Y, et al. China's CO2 peak before 2030 implied from characteristics and growth of cities. Nature Sustainability 2019;2:748-754.

[40] You K, Li Y, Cai W, Zhang L, Liu Z, Feng W, Wei Y-M. Mitigating emissions and costs through demand-side solutions in Chinese residential buildings. Nature Communications 2025;16:7358.

[41] Zou C, Ma M, Zhou N, Feng W, You K, Zhang S. Toward carbon free by 2060: A decarbonization roadmap of operational residential buildings in China. Energy 2023;277:127689.

[42] Yan R, Xiang X, Cai W, Ma M. Decarbonizing residential buildings in the developing world: Historical cases from China. Science of The Total Environment 2022;847:157679.

[43] Li H, Hong T. A semantic ontology for representing and quantifying energy flexibility of buildings. Advances in Applied Energy 2022;8:100113.

[44] Wen Y, Lau S-K, Leng J, Liu K. Sustainable underground environment integrating hybrid ventilation, photovoltaic thermal and ground source heat pump. Sustainable Cities and Society 2023;90:104383.

[45] Zhao J, Jiang Q, Dong X, Dong K, Jiang H. How does industrial structure adjustment reduce CO2 emissions? Spatial and mediation effects analysis for China. Energy Economics 2022;105:105704.

[46] Mianaei PK, Aliahmadi M, Faghri S, Ensaf M, Ghasemi A, Abdoos AA. Chance-constrained programming for optimal scheduling of combined cooling, heating, and power-based microgrid coupled with flexible technologies. Sustainable Cities and Society 2022;77:103502.

[47] Wang J, Kontar RE, Jin X, King J. Electrifying High-Efficiency Future Communities: Impact on Energy, Emissions, and Grid. Advances in Applied Energy 2022;6:100095.

[48] Yang S, Oliver Gao H, You F. Model predictive control in phase-change-material-wallboard-enhanced building energy management considering electricity price dynamics. Applied Energy 2022;326:120023.

[49] Zhang K, Prakash A, Paul L, Blum D, Alstone P, Zoellick J, et al. Model predictive control for demand flexibility: Real-world operation of a commercial building with photovoltaic and battery systems. Advances in Applied Energy 2022;7:100099.

[50] Zhao J, Jiang Q, Dong X, Dong K. Would environmental regulation improve the greenhouse gas benefits of natural gas use? A Chinese case study. Energy Economics 2020;87:104712.

[51] Malla S. An outlook of end-use energy demand based on a clean energy and technology transformation of the household sector in Nepal. Energy 2022;238:121810.

[52] Zhang B, Qiu R, Liao Q, Liang Y, Ji H, Jing R. Design and operation optimization of city-level off-grid hydro–photovoltaic complementary system. Applied Energy 2022;306:118000.

[53] Zhou S, Xu Z. Energy efficiency assessment of RCEP member states: A three-stage slack based measurement DEA with undesirable outputs. Energy 2022;253:124170.

[54] Yuan H, Ma M, Zhou N, Xie H, Ma Z, Xiang X, Ma X. Battery electric vehicle charging in China: Energy demand and emissions trends in the 2020s. Applied Energy 2024;365:123153.

[55] Wu W, Skye HM. Residential net-zero energy buildings: Review and perspective. Renewable and Sustainable Energy Reviews 2021;142:110859.

[56] Chen X, Zhao B, Shuai C, Qu S, Xu M. Global spread of water scarcity risk through trade. Resources, Conservation and Recycling 2022;187:106643.

[57] Ma M, Zhang S, Liu J, Yan R, Cai W, Zhou N, Yan J. Building floorspace and stock measurement: A review of global efforts, knowledge gaps, and research priorities. Nexus 2025;2:100075.

[58] Sun M, Liu T, Wang X, Liu T, Li M, Chen G, et al. Roles of thermal energy storage technology for carbon neutrality. Carbon Neutrality 2023;1:54.

[59] Lu M, Lai J. Review on carbon emissions of commercial buildings. Renewable and Sustainable Energy Reviews 2020;119:109545.

[60] Rivera ML, MacLean HL, McCabe B. Implications of passive energy efficiency measures on life cycle greenhouse gas emissions of high-rise residential building envelopes. Energy and Buildings 2021;249:111202.

[61] Gan L, Ren H, Cai W, Wu K, Liu Y, Liu Y. Allocation of carbon emission quotas for China's provincial public buildings based on principles of equity and efficiency. Building and Environment 2022;216:108994.

[62] Andrews A, Jain RK. Beyond Energy Efficiency: A clustering approach to embed demand flexibility into building energy benchmarking. Applied Energy 2022;327:119989.

[63] He J, Wu Y, Wu J, Li S, Liu F, Zhou J, Liao M. Towards cleaner heating production in rural areas: Identifying optimal regional renewable systems with a case in Ningxia, China. Sustainable Cities and Society 2021;75:103288.

[64] Zhao N, Zhang H, Yang X, Yan J, You F. Emerging information and communication technologies for smart energy systems and renewable transition. Advances in Applied Energy 2023;9:100125.